\documentclass[conference]{IEEEtran}

\IEEEoverridecommandlockouts

\usepackage{cite}
\usepackage{amsmath,amssymb,amsfonts}
\usepackage{algorithmic}
\usepackage{graphicx}
\usepackage{textcomp}
\usepackage{xcolor}
\usepackage{comment,multirow}
\usepackage{booktabs}
\usepackage{url}

\def\BibTeX{{\rm B\kern-.05em{\sc i\kern-.025em b}\kern-.08em
    T\kern-.1667em\lower.7ex\hbox{E}\kern-.125emX}}
\begin{document}

\title{Speech Generation Speaker Poisoning: Capability Erasure in Zero-Shot Text-to-Speech
}


\author{
\IEEEauthorblockN{
Thanathai Lertpetchpun$^{1,*}$,
Thanapat Trachu$^{2,*}$,
Sai Praneeth Karimireddy$^{2}$,
and Shrikanth Narayanan$^{1}$
}
\IEEEauthorblockA{
$^{1}$Signal Analysis and Interpretation Laboratory,
University of Southern California\\
$^{2}$Thomas Lord Department of Computer Science,
University of Southern California\\
Los Angeles, CA, USA\\
\texttt{lertpetc@usc.edu}\\
$^{*}$Equal contribution
}
}

\maketitle

\begin{abstract}
Recent zero-shot Text-to-Speech (TTS) systems can clone previously unseen voices from only a few seconds of audio. We formulate \emph{Speech Generation Speaker Poisoning} (SGSP), a task that seeks to prevent a model from synthesizing targeted speaker identities while maintaining performance on all other speakers. Unlike conventional machine unlearning, removing training examples is insufficient because modern zero-shot TTS systems can reconstruct identities through learned speaker representations and strong generalization capabilities. We evaluate both inference-time filtering and parameter-modification approaches across settings involving 1, 15, and 100 forget speakers, focusing primarily on speakers seen during training, which we show are harder to suppress than unseen speakers. To characterize the trade-off between utility and privacy, we introduce an evaluation framework based on AUC analysis and a proposed metric, Forget Set Similarity (FSSIM). Our results demonstrate effective speaker suppression for up to 15 forget speakers while revealing that worst-case identity leakage (Max-FSSIM) remains unresolved at multi-speaker scale — establishing this as an open challenge for future work. Together, our work establishes targeted speaker poisoning as a task for zero-shot TTS and, using StyleTTS2 as an initial testbed, provides methods and evaluation protocols for future research, with code and model weights\footnote{\url{https://github.com/NewGamezzz/TargetedSpeakerPoisoning}}.
\end{abstract}

\begin{IEEEkeywords}
Text-To-Speech Model, Poisoning Generation Model, Speech Privacy, Speech Generation
\end{IEEEkeywords}

\section{Introduction}

The rapid evolution of generative AI has advanced Text-to-Speech (TTS) systems far beyond conventional speech synthesis, enabling high-fidelity voice cloning from audio prompts as short as three seconds. Recent models~\cite{styletts2,valle,voicebox} demonstrate remarkable expressiveness and speaker similarity. However, this capability also introduces serious real-world risks. Studies~\cite{barrington2025peoplepoorlyequippeddetect} show that human listeners struggle to distinguish AI-generated speech from genuine recordings, making these systems susceptible to misuse, including impersonation of public figures~\cite{rosen2024fake} and the dissemination of misinformation~\cite{destefano2023kidnapping}. 

These risks have motivated a parallel line of work on voice anonymization, which protects speaker identity in recorded speech by transforming audio so that the speaker cannot be identified~\cite{tomashenko2022voiceprivacy, tomashenko2024voiceprivacy}. Our work addresses a complementary but distinct use case: rather than protecting the identity of a speaker in a recording, we aim to prevent a trained generative model from reproducing a specific target identity and guard against impersonation and misuse at the level of the model itself.

A closely related research thread is machine unlearning, which aims to selectively remove particular knowledge from a trained model~\cite{wang2024machineunlearningcomprehensivesurvey,9519428}. However, this paradigm is not directly applicable to modern zero-shot TTS systems. Recent models~\cite{du2025cosyvoice, ju2024naturalspeech, chen2025f5} can clone previously ``unseen" speakers from only a few seconds of reference audio, meaning speaker identity is not associated with training examples but emerges from a learned representation space that supports strong generalization. As a result, approximating the parameters of an unlearned model does not necessarily prevent the reconstruction of their identities. Addressing this challenge requires a new problem formulation, along with relevant baselines and an evaluation framework that explicitly accounts for zero-shot speaker cloning. Concurrent work has begun to explore this space: \cite{kim2026continual} extend speaker identity unlearning to a continual setting where removal requests arrive sequentially.

This problem is conceptually related to model poisoning~\cite{10646851}, where the objective is to condition a model against performing certain behaviors. We formalize this goal as \emph{Speech Generation Speaker Poisoning} (SGSP), a capability-erasure task for zero-shot TTS systems. Following the terminology of machine unlearning, we define a forget set consisting of speaker identities that the model should not synthesize and a retain set containing speakers whose synthesis capability must be preserved. Our goal is to modify the model such that, when prompted with samples from the forget set, it fails to reproduce the corresponding speaker identity while maintaining performance on the retain set.

A straightforward approach to this problem is to apply pre-processing strategies that replace reference speakers from the forget set with speakers from the retain set by using a certain criteria such as speaker similarity. However, such external filtering approaches are bypassable by an adversary with direct model access.~\cite{9519486, 10.1145/3376897.3377856}. A concurrent approach~\cite{lee2026erasing} addresses suppression at inference time through training-free representation shifting, without modifying model parameters; however, such methods rely on server-side access control and remain bypassable when weights are distributed publicly. Therefore, we focus on methods that directly modify the internal model parameters to achieve targeted speaker suppression.

To this end, we build upon the Teacher-Guided Poisoning (TGP) framework \cite{unlearn_speaker}, originally proposed for VoiceBox~\cite{voicebox}, and adapt it to StyleTTS2~\cite{styletts2}. TGP follows a knowledge distillation paradigm \cite{hinton2015distillingknowledgeneuralnetwork}, where targets from a teacher model guide the speaker poisoning process. We additionally propose Encoder-Guided Poisoning (EGP), which directly learns speaker representations from ground-truth data. We further incorporate a contrastive objective to explicitly suppress forgotten identities. Together, these components form an efficient framework for targeted speaker suppression in zero-shot TTS.

Systematic evaluation protocols remain underdeveloped in SGSP. Prior work \cite{unlearn_speaker} relies primarily on raw speaker similarity to assess the dissimilarity between the synthesized speech and the targeted forget speakers. However, this approach is difficult to interpret, as the effective range depends heavily on the employed speaker verification model~\cite{Ferrer_2022}. Moreover, average similarity alone does not reflect whether the similarity distributions of the retain and forget sets are well separated. To address these limitations, we introduce a distribution-aware evaluation framework based on Area Under the Curve (AUC), together with a new privacy metric, Forget Set Similarity (FSSIM), which measures similarity between generated samples and all other speakers in the forget set.

The main contributions of this work are as follows: 1)~We formulate \emph{Speech Generation Speaker Poisoning} (SGSP) as a capability-erasure problem for generative speech models, defining the forget and retain sets, and empirically validate that suppressing seen (in-domain) speakers is a stricter benchmark than suppressing unseen speakers. 2)~We establish naive pre-processing baselines to highlight the limitations of external filtering approaches. 3)~We adapt Teacher-Guided Poisoning (TGP) to StyleTTS2 and further propose Encoder-Guided Poisoning (EGP) with a triplet-loss objective for explicit identity suppression. 4)~We develop a comprehensive evaluation framework, including distribution-level AUC analysis and a new privacy metric, FSSIM, to systematically assess the effectiveness of speaker poisoning. Training code, baselines, model weights, and the evaluation framework will be released upon acceptance.

\vspace{-2mm}
\section{Problem Formulation}
Let $\mathcal{S}$ denote the set of speakers that the TTS model is originally trained to generate. Our objective is to manipulate the model such that it cannot generate speech corresponding to any speakers in a forget set $\mathcal{F}\subseteq \mathcal{S}$, while preserving the ability to correctly synthesize speech for all other speakers in the retain set $\mathcal{R}=\mathcal{S}\setminus \mathcal{F}$.

We primarily construct $\mathcal{F}$ from speakers observed during training rather than from unseen test speakers, motivated by the following hypothesis. In modern zero-shot TTS systems, speaker identities emerge from a representation space learned during training; training speakers are therefore the identities most directly encoded by the model, while unseen speakers are synthesized only through generalization from this representation space. Recent work \cite{valle, jeon2024enhancing} shows that models reproduce in-domain speakers with higher fidelity than out-of-domain speakers, suggesting that in-domain identities should be harder to suppress than unseen ones. If this is the case, seen speakers pose the stricter test for capability erasure, since the model represents them strongly, so successfully suppressing them is real evidence the method works, unlike succeeding only on weakly represented, unseen speakers.

We test this hypothesis by repeating our core experiments on both seen and unseen speakers under matched conditions for the 1- and 15-speaker forget settings (Section~\ref{sec:unseen}). Following this formulation, we evaluate our models across three distinct settings: 1, 15, and 100 speakers in the forget set $\mathcal{F}$, with seen speakers as our main focus across all three settings, and unseen speakers included as a complementary test at the 1- and 15-speaker scales.

\begin{figure*}[ht!]
    \centering
    \includegraphics[width=.90\linewidth]{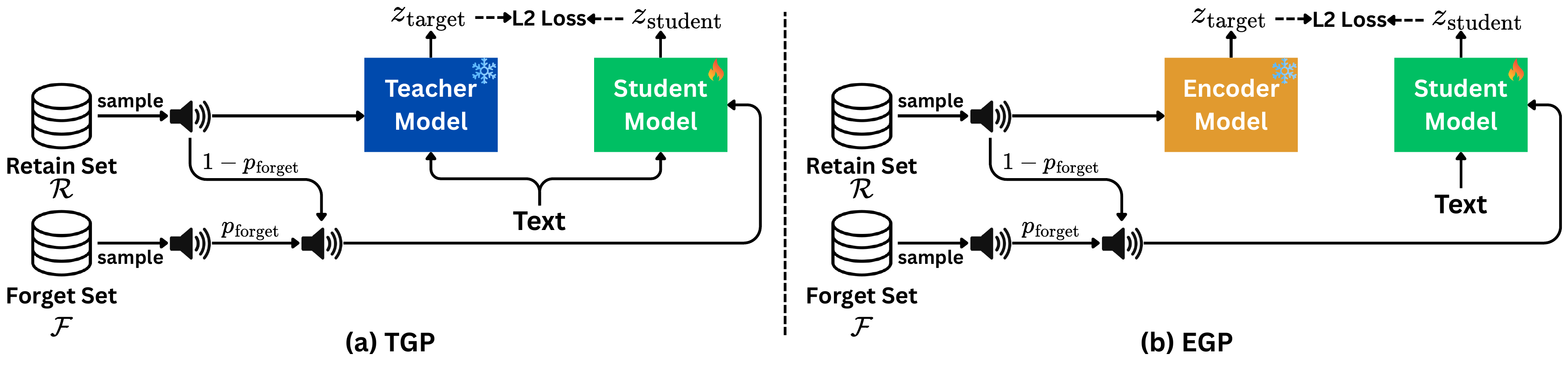}
    \caption{Overview of the proposed poisoning framework. A retain-speaker target is paired with a reference speaker that is replaced by a forget speaker with probability $p_{\text{forget}}$. The student model learns to map forget-speaker references toward retain-speaker representations. TGP uses targets generated by a frozen pretrained diffusion model, whereas EGP uses targets extracted directly from the StyleTTS2 style encoder.}
    \label{fig:overview}
    \vspace{-5mm}
\end{figure*}

\section{SGSP Baselines}

\subsection{StyleTTS2 Background}
StyleTTS2~\cite{styletts2} is a state-of-the-art TTS model composed of six modules, which can be grouped into three components: a speech generation system (text encoder, style encoder, and speech decoder), a TTS prediction system (duration and prosody predictors), and a diffusion sampler (diffusion model). The training process involves several loss functions. The main losses include the mel-spectrogram reconstruction loss (L2), an adversarial loss similar to GAN training, and the diffusion loss, which is an L2 loss between the style-encoder output from the first stage and the output of the diffusion model. As our study focuses on manipulating speaker identity, we fine-tune only the diffusion module using the diffusion loss. This ensures that the speaker poisoning process specifically targets the speaker identity representation, preventing degradation of the other abilities.

\subsection{Naïve Baselines}
We introduce two naïve algorithm to serve as baselines for SGSP. Given a reference utterance for TTS conditioning, these methods first determine whether the reference speaker belongs to the forget set or the retain set. If the speaker is identified as belonging to the forget set, the reference utterance is replaced with an alternative utterance from the retain set.

\vspace{1mm}
\noindent \textbf{Pretrained + Speaker Filtering:} We utilize WavLM \cite{wavlm} to extract speaker embeddings and compute the cosine similarity between the given reference and all utterances in the forget set $\mathcal{F}$. Following the default threshold of Wavlm from Hugging Face\footnote{\url{https://huggingface.co/microsoft/wavlm-base-plus-sv}}, the reference speaker is classified as belonging to the forget set if the cosine similarity with any utterance in $\mathcal{F}$ exceeds 0.86. It is then dynamically replaced by iteratively sampling from the retain set $\mathcal{R}$ until a new reference with a similarity strictly below 0.86 is found. 

\vspace{1mm}
\noindent \textbf{Pretrained + Ground Truth Filtering:} This baseline follows the exact same replacement procedure but assumes perfect, ground-truth knowledge of whether the reference utterance belongs to $\mathcal{F}$ or $\mathcal{R}$, thereby bypassing the initial speaker filtering step.

\subsection{Parameter-Modifying Baselines}

\subsubsection{Speaker Poisoning Objective}

Let $G(x,s)$ denote the speaker-conditioned generation process of a TTS model, where $x$ is the input transcript and $s$ is a speaker representation extracted from a reference utterance. Our objective is to modify the model such that references belonging to the forget set $\mathcal{F}$ no longer reproduce their original speaker identities. Instead, when conditioned on a forget speaker $s_f \in \mathcal{F}$, the model should generate speech corresponding to a randomly sampled retain speaker $s_r \in \mathcal{R}$ while preserving normal behavior for retain speakers. During fine-tuning, we sample retain speakers as usual and replace the retain speaker reference with a forget speaker reference with probability $p_{\text{forget}}$. The target speaker identity remains unchanged, encouraging the model to associate forget references with retain identities. Formally, we seek to learn a modified generator $\hat{G}$ satisfying
\begin{align}
\hat{G}(x,s_f) &\approx G(x,s_r), \quad s_f \in \mathcal{F}, \; s_r \in \mathcal{R}, \\
\hat{G}(x,s_r) &\approx G(x,s_r), \quad s_r \in \mathcal{R}
\end{align}
To achieve this objective, we optimize the student model to match a target representation corresponding to a retain speaker while conditioning the student on a forget speaker. Let $z_{\text{student}}$ denote the diffusion output of the student model and $z_{\text{target}}$ denote the target representation. The objective is defined as
\begin{align}
\label{eq:poison}
\mathcal{L}_{\text{poison}}
=
\left\|
z_{\text{student}} - z_{\text{target}}
\right\|_2^2
\end{align}
The choice of $z_{\text{target}}$ distinguishes the two variants explored in this work. Teacher-Guided Poisoning (TGP) uses outputs generated by a teacher model as targets, whereas Encoder-Guided Poisoning (EGP) directly uses speaker representations extracted from the StyleTTS2 style encoder.

\vspace{1mm}
\noindent \textbf{Teacher-Guided Poisoning (TGP).}
TGP adopts a teacher-student framework in which the original pretrained StyleTTS2 diffusion model serves as a frozen teacher. For each training sample, the teacher is conditioned on a randomly sampled retain speaker to produce the target representation $z_{\text{target}}$. The student model, initialized from the same pretrained diffusion checkpoint, is then optimized using Eq.~(\ref{eq:poison}) to match the teacher output.

\vspace{1mm}
\noindent \textbf{Encoder-Guided Poisoning (EGP).}
EGP uses the same procedure as TGP, but the target is the speaker representation from the StyleTTS2 style encoder instead of the teacher's output.

\subsubsection{Contrastive Learning}
We employ contrastive learning to explicitly penalize outputs that remain similar to speaker identities in the forget set $\mathcal{F}$. Specifically, we use a triplet loss~\cite{triplet_loss} that encourages the student representation $z_{\text{student}}$ to remain close to the target retain-speaker representation $z_{\text{target}}$ while increasing its distance from a negative representation $z_{\text{neg}}$ extracted from a speaker in $\mathcal{F}$. To preserve utility, this penalty is applied only when the model is conditioned on a forget speaker:
\begin{equation}
\mathcal{L}_{\text{triplet}} = \left[
\|z_{\text{student}}-z_{\text{target}}\|_2^2
-\|z_{\text{student}}-z_{\text{neg}}\|_2^2 + \beta
\right]_+
\label{eq:triplet}
\end{equation}
where $\beta$ is a margin that enforces a minimum separation between the target retain-speaker representation and the negative forget-speaker representation.

The final training objective combines the poisoning loss in Eq.~(\ref{eq:poison}) and the contrastive loss in Eq.~(\ref{eq:triplet}):
\begin{align}
\mathcal{L}
&=
\mathcal{L}_{\text{poison}}
+
\lambda \mathcal{L}_{\text{triplet}} 
\label{eq:total_loss}
\end{align}
where $\lambda$ controls the strength of the contrastive regularization.

\section{Evaluation Metrics}

\subsection{Utility Metrics}
To ensure the model maintains high-quality synthesis, we evaluate speech intelligibility, naturalness, and speaker consistency. Word Error Rate (WER) is computed using the Whisper-medium ASR model \cite{whisper} on utterances shorter than 30 seconds. We utilize UTMOS \cite{utmos} as an proxy to evaluates the perceived naturalness of synthesized speech on a scale from 1 (worst quality) to 5 (best quality). To measure identity preservation, we compute the cosine similarity between embeddings of reference and synthesized speech. Embeddings are extracted using the \texttt{wavlm-base-plus-sv} model\cite{wavlm}, where values closer to 1 signify higher speaker consistency.

\subsection{Privacy Metrics}

We evaluate privacy preservation under two progressively stricter dissimilarity conditions:

\vspace{1mm}
\noindent \textbf{Easy Condition: Prompt-Output Dissimilarity.}
Under this condition, we compute \textbf{Speaker Similarity (SSIM)} as the cosine similarity between a synthesized utterance and its reference prompt. For retained speakers, high SSIM indicates successful identity preservation, whereas for forget speakers, low SSIM indicates effective suppression. To provide a distribution-level assessment, we additionally compute the \textbf{Area Under the Curve (AUC)} between the retain and forget SSIM distributions. An AUC of 0.5 indicates no separation, whereas an AUC of 1.0 indicates perfect separation. This condition evaluates whether the model preserves retained identities while avoiding reproduction of forgotten identities.

\vspace{1mm}
\noindent \textbf{Strong Condition: Dissimilarity to All Other Forget Speakers.}
While pairwise similarity metrics evaluate whether a synthesized utterance resembles its corresponding forget speaker, they do not guarantee dissimilarity from the remaining speakers in the forget set. To enforce a stronger notion of identity suppression, we introduce \textbf{Forget Set Similarity (FSSIM)}, which measures the similarity between each synthesized utterance and all \textit{other} speakers in the forget set (excluding the speaker used as its reference). We summarize these similarities using two complementary metrics: \textbf{Avg-FSSIM}, which captures the average similarity across the remaining forget speakers, and \textbf{Max-FSSIM}, which captures the highest similarity to any of them. The latter provides a stringent worst-case measure, ensuring that synthesized speech does not inadvertently resemble any other individual whose identity should be forgotten.

Let $\mathcal{F}$ denote the set of forget speakers, $\mathcal{S}$ the set of synthesized utterances generated using a reference from speaker $s_i \in \mathcal{F}$, $\bar{e}_s$ the average embedding of speaker $s$, $\phi(u_i)$ the speaker embedding of synthesized utterance $u_i$, and $\mathrm{sim}(\cdot,\cdot)$ a cosine similarity function. We define Avg-FSSIM as
\begin{align}
\text{Avg-FSSIM}
&= \mathbb{E}_{u_i \in \mathcal{S}}
\!\left[
\tfrac{1}{|\mathcal{F}|-1}
\sum_{s \in \mathcal{F} \setminus \{s_i\}}
\mathrm{sim}\!\left(\phi(u_i), \bar{e}_{s}\right)
\right],
\end{align}
which measures how similar each synthesized utterance is, on average, to the forget speakers other than the one used as its reference. To capture the strongest remaining identity leakage, we additionally define Max-FSSIM as
\begin{align}
\text{Max-FSSIM}
&=
\mathbb{E}_{u_i \in \mathcal{S}}
\left[
\max_{s \in \mathcal{F} \setminus \{s_i\}}
\mathrm{sim}\!\left(\phi(u_i),\, \bar{e}_{s}\right)
\right].
\end{align}
Lower values of Avg-FSSIM and Max-FSSIM indicate stronger forgetting, with Max-FSSIM providing a stricter assessment by focusing on the most similar remaining forget speaker for each synthesized utterance. Note that both metrics are undefined when $|\mathcal{F}|=1$, as there is no other forget speaker to compare against.

\section{Experimental Setup}
\label{sec:data_creation}
\subsection{Dataset}
We evaluate our methods using the LibriTTS-R \cite{libritts} dataset. The data construction follows two steps:

\vspace{0.5mm}
\noindent \textbf{Speaker and Utterance Partitioning for Seen Settings.}
Combining the \texttt{train-clean-100} and \texttt{train-clean-360} subsets, we randomly sample $n$ (1, 15, or 100) seen speakers to form the \textit{forget set} and assign the remaining speakers to the \textit{retain set}. For each speaker, utterances are split 90:10 into training and testing partitions, yielding four disjoint subsets: \textit{forget train}, \textit{retain train}, \textit{forget test}, and \textit{retain test}. During finetuning, \textit{retain train} is used to construct retain-speaker targets, while \textit{forget train} provides the speaker references to be poisoned. In evaluation, both \textit{retain test} and \textit{forget test} are used to assess utility through WER and UTMOS. Speaker similarity on \textit{retain test} and \textit{forget test} is used to compute SSIM$_{\mathcal{R}}$ and SSIM$_{\mathcal{F}}$, respectively. These similarity distributions are further used to compute AUC. Finally, \textit{forget test} is used to compute FSSIM, which evaluates whether synthesized samples remain similar to any other speakers in the forget set.

\vspace{1mm}
\noindent \textbf{Transcript Pairing.}
During finetuning, each utterance in \textit{retain train} is paired with a randomly sampled transcript from the \texttt{train-clean-100} and \texttt{train-clean-360} subsets while utterances in \textit{forget train} are not assigned fixed a transcript pair; instead, they are dynamically paired with retain-set transcripts during training. During evaluation, each utterance in \textit{retain test} and \textit{forget test} is similarly paired with a randomly sampled transcript from the \texttt{test-clean} subset. This procedure ensures a consistent number of inference (positive and negative) trials for both retain and forget speakers by selecting the same number of transcript for pairing.

\vspace{1mm}
\noindent \textbf{Unseen Speaker Pool.}
To evaluate generalization to speakers not seen during training, we additionally construct forget/retain partitions following the same procedure using the \texttt{dev-clean} and \texttt{dev-other} subsets, which together contain only 73 speakers. As this pool is not large enough to support a 100-speaker forget set, we restrict the unseen-speaker evaluation to the 1- and 15-speaker settings.

\begin{table*}[t]
    \centering
    \scriptsize
    \vspace{-3mm}
    \caption{Model performance evaluations for the \textbf{15} \textbf{unseen} speaker setting. $\mathcal{R}$ and $\mathcal{F}$ represent the retain and forget sets, respectively. PT stands for Pretrained, SF and GTF denotes speaker filtering and ground truth filtering, respectively. T is Triplet loss.}    
    \vspace{-2mm}
    \label{tab:15_unseen_speakers}
    \setlength{\tabcolsep}{12pt}
    \begin{tabular}{l cc cc cc c cc}
        \toprule
        \multirow{2}{*}{\textbf{Method}}
        & \multicolumn{2}{c}{\textbf{WER}}
        & \multicolumn{2}{c}{\textbf{MOS}}
        & \multicolumn{2}{c}{\textbf{SSIM}}
        & \multirow{2}{*}{\textbf{AUC}$\uparrow$}
        & \multicolumn{2}{c}{\textbf{FSSIM}}\\
        \cmidrule(lr){2-3} \cmidrule(lr){4-5} \cmidrule(lr){6-7} \cmidrule(lr){9-10}
        & $\mathcal{R} \downarrow$ & $\mathcal{F} \downarrow$
        & $\mathcal{R} \uparrow$ & $\mathcal{F} \uparrow$
        & $\mathcal{R} \uparrow$ & $\mathcal{F} \downarrow$
        & & Avg $\downarrow$ & Max $\downarrow$
        \\
        \midrule
        PT        & 2.61 & \textbf{2.66} & \textbf{4.30} & \textbf{4.32} & \textbf{0.89} & 0.90 & 0.46 & 0.68 & 0.91 \\
        PT + SF   & 2.73 & 2.73 & 4.29 & 4.29 & 0.68 & 0.64 & 0.57 & 0.68 & 0.90 \\
        PT + GTF  & \textbf{2.58} & 2.75 & \textbf{4.30} & 4.29 & 0.89 & 0.68 & 0.87 & 0.68 & 0.91 \\
        \midrule
        TGP           & 2.85 & \textbf{2.66} & 4.26 & 4.28 & 0.84 & 0.73 & 0.76 & 0.70 & 0.88 \\
        TGP + T & 2.78 & 2.67 & 4.26 & 4.28 & 0.85 & 0.74 & 0.75 & 0.69 & 0.87 \\
        EGP           & 2.76 & 2.83 & 4.28 & 4.29 & 0.88 & 0.74 & 0.78 & 0.72 & 0.91 \\
        EGP + T & 3.01 & 18.17 & 4.28 & 3.29 & 0.85 & \textbf{0.65} & \textbf{0.81} & \textbf{0.62} & \textbf{0.81} \\
        \bottomrule
    \end{tabular}
    \vspace{-5mm}
\end{table*}

\vspace{-1mm}
\subsection{Model}
In this work, we employ StyleTTS2 \cite{styletts2} as a backbone model. We utilize the official pretrained checkpoint trained on the LibriTTS corpus\footnote{https://github.com/yl4579/StyleTTS2}. Crucially, during the fine-tuning, we freeze all major components (e.g., the text encoder, decoders, and discriminators) and fine-tune only the diffusion module. The model is fine-tuned for 60,000 steps using the AdamW optimizer with a learning rate of $1 \times 10^{-4}$. For the triplet-loss configuration, we set $\lambda$ to 1.0 with a margin $\beta$ of 0.3. The forget ratio $p_\text{forget}$ is set to 0.5 across all experiments.

\section{Results and Discussion}
We first test our hypothesis that seen speakers are harder to suppress than unseen ones, comparing both under matched 1- and 15-speaker settings (Section~\ref{sec:unseen}). We then analyze the seen-speaker setting in depth across all three forget-set sizes (1, 15, 100), examining single-speaker (Section~\ref{sec:single_seen}) and multi-speaker scalability (Section~\ref{sec:multi_seen}).

\begin{table}[t]
    \centering
    \scriptsize
    \caption{Performance evaluations for the \textbf{1 unseen} speaker setting.}

    \begin{tabular}{l cc cc cc c}
        \toprule
        \multirow{2}{*}{\textbf{Method}} & \multicolumn{2}{c}{\textbf{WER}} & \multicolumn{2}{c}{\textbf{MOS}} & \multicolumn{2}{c}{\textbf{SSIM}} & \multirow{2}{*}{\textbf{AUC} $\uparrow$}\\
        \cmidrule(lr){2-3} \cmidrule(lr){4-5} \cmidrule(lr){6-7}
         & $\mathcal{R} \downarrow$ & $\mathcal{F} \downarrow$ & $\mathcal{R} \uparrow$ & $\mathcal{F} \uparrow$ & $\mathcal{R} \uparrow$ & $\mathcal{F} \downarrow$ &   \\
         \midrule
        PT        & 3.39 & 2.70 & 4.30 & \textbf{4.35} & 0.87 & 0.85 & 0.59 \\
        PT + SF   & 2.64 & \textbf{2.64} & 4.30 & 4.30 & 0.67 & 0.61 & 0.60 \\
        PT + GTF  & \textbf{2.58} & \textbf{2.64} & 4.30 & 4.30 & 0.87 & 0.61 & 0.93 \\
        \midrule
        TGP       & 2.88 & 2.95 & 4.12 & 4.13 & 0.84 & 0.64 & 0.90 \\
        TGP + T   & 2.82 & 3.06 & 4.17 & 4.16 & 0.84 & 0.60 & 0.92 \\
        EGP       & 2.79 & 2.79 & \textbf{4.31} & 4.31 & \textbf{0.88} & 0.71 & 0.88 \\
        EGP + T   & 2.75 & 9.15 & 4.27 & 2.29 & 0.87 & \textbf{0.44} & \textbf{0.99} \\
        \bottomrule
    \end{tabular}
    \vspace{-5mm}
    \label{tab:1_unseen_speaker}
\end{table}

\subsection{Seen-vs-Unseen Hypothesis}
\label{sec:unseen}
Comparing Tables~\ref{tab:1_speaker}/\ref{tab:15_100_speakers} (seen) against Tables~\ref{tab:1_unseen_speaker}/\ref{tab:15_unseen_speakers} (unseen) shows that, for the best-performing method in each setting, unseen speakers achieve consistently higher AUC than seen speakers (0.99 vs.\ 0.95 at 1 speaker; 0.81 vs.\ 0.77 at 15 speakers), indicating cleaner separation between retain and forget distributions. This is consistent with unseen identities being synthesized through generalization rather than direct optimization, and is consistent with our hypothesis that seen speakers present the stricter setting. We therefore focus on seen speakers for the remainder of this work.

\subsection{Single Seen-Speaker Setting}
\label{sec:single_seen}
\noindent \textbf{Performance Overview.} As shown in Table~\ref{tab:1_speaker}, most baselines maintain high utility (WER and MOS) comparable to the pretrained model. However, PT + SF provides little privacy benefit, indicating that simple similarity-based filtering using the default threshold (0.86) is insufficient in the single speaker setting. In contrast, PT + GTF demonstrates that strong privacy and high utility can coexist under an ideal filtering mechanism. Among the parameter-modifying approaches, TGP and EGP preserve utility while reducing speaker similarity in $\mathcal{F}$. EGP+Triplet achieves the strongest privacy performance, albeit with a noticeable degradation in utility. 

\vspace{0.5mm}
\noindent\textbf{Comparison: EGP vs. TGP.} EGP consistently outperforms TGP variants. We believe that this is because the teacher and student share identical model capacities; TGP's reliance on teacher-generated ground truth introduces unnecessary generative noise. As demonstrated in \cite{stanton2021does}, when the student model's size matches the teacher model, it struggles to further improve performance through distillation. EGP circumvents this by targeting the original style encoder representations, providing a cleaner target and superior performance.

\begin{table}[t]
    \centering
    \scriptsize

    \caption{Performance evaluations for the \textbf{1 seen} speaker setting.}
    \begin{tabular}{l cc cc cc c}
        \toprule
        \multirow{2}{*}{\textbf{Method}} & \multicolumn{2}{c}{\textbf{WER}} & \multicolumn{2}{c}{\textbf{MOS}} & \multicolumn{2}{c}{\textbf{SSIM}} & \multirow{2}{*}{\textbf{AUC} $\uparrow$}\\
        \cmidrule(lr){2-3} \cmidrule(lr){4-5} \cmidrule(lr){6-7}
         & $\mathcal{R} \downarrow$ & $\mathcal{F} \downarrow$ & $\mathcal{R} \uparrow$ & $\mathcal{F} \uparrow$ & $\mathcal{R} \uparrow$ & $\mathcal{F} \downarrow$ &   \\
         \midrule
        PT &  2.75 & 2.72 & 4.29 & 4.27 & 0.87 & 0.88 & 0.47\\
        PT + SF &  2.47 & 2.57 & 4.29 & 4.29 & 0.63 & 0.63 & 0.50\\
        PT + GTF &  2.75 & 2.57 & 4.29 & 4.29 & 0.87 & 0.63 & 0.90\\
        \midrule
        TGP &  2.74 & \textbf{2.62} & 4.34 & 4.34 & 0.85 & 0.77 & 0.71\\
        TGP + T & \textbf{2.69} & 2.85 & \textbf{4.36} & \textbf{4.36} & 0.85 & 0.74 & 0.74\\
        EGP & 2.90 & 3.00 & 4.28 & 4.29 & \textbf{0.88} & 0.77 & 0.79\\
        EGP + T & 2.89 & 7.86 & 4.27 & 3.38 & 0.87 & \textbf{0.48} & \textbf{0.95}\\
        \bottomrule
    \end{tabular}
    \vspace{-5mm}
    \label{tab:1_speaker}
\end{table}

\subsection{Multiple Seen-Speakers Setting}
\begin{figure*}
    \centering
    \includegraphics[width=0.8\linewidth]{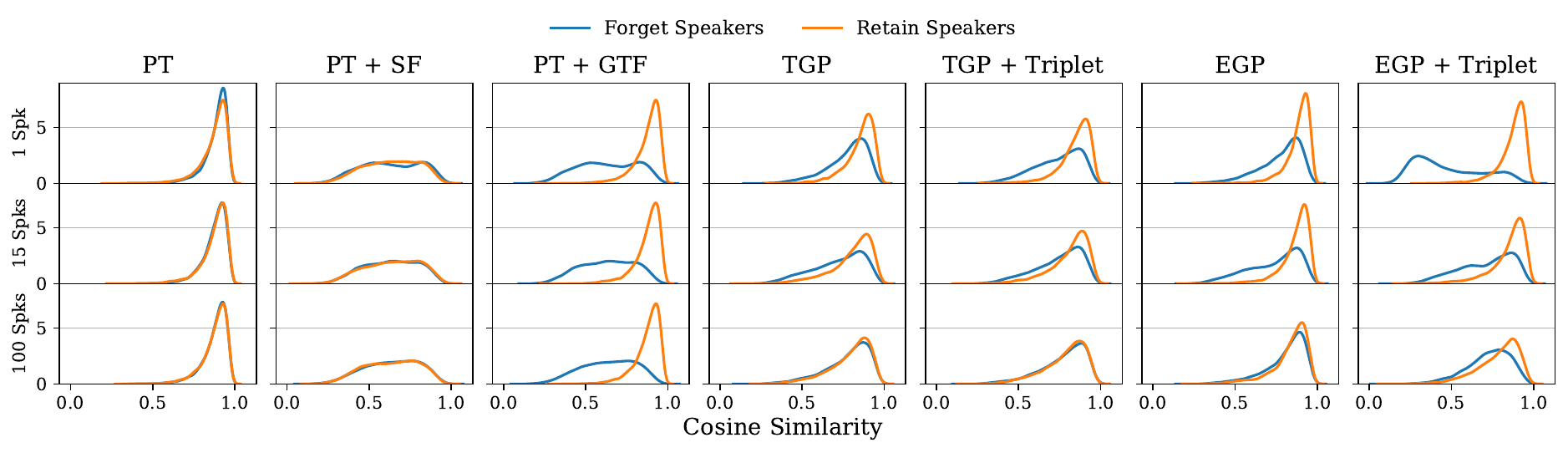}
    \vspace{-5mm}
    \caption{Cosine similarity distributions for the retained speaker set (in orange) and for the forget set (in blue). Rows correspond to different numbers of speakers (1, 15, or 100) in the forget set, whereas columns correspond to different system configurations.}
    \label{fig:auc_comparison}
\end{figure*}
\begin{table*}[t]
    \centering
    \scriptsize
    \vspace{-3mm}
    \caption{Model performance evaluations for the \textbf{15} and \textbf{100} \textbf{seen} speaker settings.}
    \vspace{-2mm}
    \setlength{\tabcolsep}{12pt}
    \begin{tabular}{l cc cc cc c cc}
        \toprule
        \multirow{2}{*}{\textbf{Method}} 
        & \multicolumn{2}{c}{\textbf{WER}}
        & \multicolumn{2}{c}{\textbf{MOS}}
        & \multicolumn{2}{c}{\textbf{SSIM}} 
        & \multirow{2}{*}{\textbf{AUC}$\uparrow$}
        & \multicolumn{2}{c}{\textbf{FSSIM}} \\
        \cmidrule(lr){2-3} \cmidrule(lr){4-5} \cmidrule(lr){6-7} \cmidrule(lr){9-10}
        & $\mathcal{R} \downarrow$ & $\mathcal{F} \downarrow$
        & $\mathcal{R} \uparrow$ & $\mathcal{F} \uparrow$
        & $\mathcal{R} \uparrow$ & $\mathcal{F} \downarrow$ 
        & & Avg $\downarrow$ & Max $\downarrow$ \\
        \midrule
        \multicolumn{10}{c}{\textbf{15 Seen Speakers Setting}} \\
        \midrule
        PT          & 2.72 & 2.77 & 4.29 & 4.28 & 0.87 & 0.88 & 0.51 & 0.67 & 0.91 \\
        PT + SF     & 2.53 & 2.56 & 4.29 & 4.29 & 0.64 & 0.63 & 0.51 & 0.70 & 0.91 \\
        PT + GTF    & 2.72 & 2.56 & 4.29 & 4.29 & 0.87 & 0.63 & 0.91 & 0.70 & 0.91 \\
        \midrule
        TGP             & 2.77 & 2.74 & 4.34 & 4.34 & 0.81 & 0.72 & 0.66 & 0.71 & \textbf{0.91} \\
        TGP + T   & 2.70 & 2.76 & 4.31 & 4.32 & 0.81 & 0.74 & 0.64 & 0.72 & \textbf{0.91} \\
        EGP             & 2.84 & 2.85 & 4.30 & 4.30 & \textbf{0.87} & 0.74 & \textbf{0.77} & 0.70 & \textbf{0.91} \\
        EGP + T& 2.84 & 3.94 & 4.31 & 4.01 & 0.85 & \textbf{0.71} & 0.76 & \textbf{0.69} & \textbf{0.91} \\
        \midrule
        \multicolumn{10}{c}{\textbf{100 Seen Speakers Setting}} \\
        \midrule
        PT          & 2.85 & 2.80 & 4.29 & 4.30 & 0.87 & 0.88 & 0.49 & 0.70 & 0.94 \\
        PT + SF     & 2.54 & 2.54 & 4.29 & 4.30 & 0.63 & 0.64 & 0.49 & 0.70 & 0.94 \\
        PT + GTF    & 2.85 & 2.54 & 4.29 & 4.30 & 0.87 & 0.64 & 0.91 & 0.70 & 0.94 \\
        \midrule
        TGP             & 2.72 & 2.80 & 4.32 & 4.33 & 0.79 & 0.78 & 0.53 & 0.74 & 0.95 \\
        TGP + T   & 2.63 & 2.76 & 4.36 & 4.36 & 0.78 & 0.77 & 0.51 & 0.74 & 0.95 \\
        EGP             & 2.97 & 2.91 & 4.27 & 4.28 & \textbf{0.83} & 0.81 & 0.57 & 0.72 & 0.94 \\
        EGP + T   & 3.67 & 5.25 & 4.14 & 3.74 & 0.78 & \textbf{0.72} & \textbf{0.64} & \textbf{0.71} & \textbf{0.91} \\
        \bottomrule
    \end{tabular}
    \vspace{-5mm}
    \label{tab:15_100_speakers}
\end{table*}
\label{sec:multi_seen}

\noindent \textbf{Distributional Analysis of AUC.}
Figure~\ref{fig:auc_comparison} shows the cosine similarity distributions of the retain and forget sets. Pretrained and Pretrained + SF exhibit substantial overlap, yielding AUC values near 0.5, whereas Pretrained + GTF and EGP + Triplet produce well-separated distributions and correspondingly high AUC scores. In most settings, incorporating triplet loss generally shifts the forget-set distribution toward lower similarity values, further increasing distributional separation and AUC. Overall, separation between retain and forget distributions narrows monotonically as the forget set grows, reflecting the increasing difficulty of isolating individual identities within a denser learned speaker space.

\vspace{0.5mm}
\noindent \textbf{Scalability Challenges in Multi-Speaker SGSP.} When scaling to multiple forget speakers (Table~\ref{tab:15_100_speakers}), all methods face increased difficulty suppressing target compared to the single-speaker setting. For 15 speakers, parameter-modifying approaches maintain a similarity gap between the retain ($\mathcal{R}$) and forget ($\mathcal{F}$) sets. However, in the 100-speaker setting, this distinction largely collapses as shown in Figure~\ref{fig:auc_comparison}. Despite the deterioration in privacy performance, utility remains largely stable for TGP, TGP + Triplet, and EGP across all settings. In contrast, EGP + Triplet exhibits the strongest privacy-utility trade-off, with noticeably worse WER and MOS.

\vspace{0.5mm}
\noindent \textbf{Strong Privacy Condition.}
Under the strong privacy condition, which requires suppressing all identities in the forget set, the results reveal a clear distinction between average-case and worst-case behavior. While the Avg-FSSIM remains below the speaker-verification threshold of 0.86, the Max-FSSIM remains high across both multi-speaker settings. This suggests that the model continues to generate samples that are highly similar to at least one speaker in the forget set, even when average similarity is low.

\vspace{0.5mm}
\noindent \textbf{The Privacy-Utility Trade-off.} Triplet loss also introduces a separate privacy-utility trade-off: EGP+Triplet maximizes privacy but degrades $\mathcal{F}$ utility compared to base EGP. This trade-off holds across all forget-set sizes, since the loss pushes each embedding away from $\mathcal{F}$ regardless of how many other forget speakers are present. We conjecture that this push forces embeddings off the manifold of natural speaker styles \cite{arvanitidis2018latent}, giving the decoder an out-of-distribution vector that degrades output quality, consistent with the catastrophic collapse observed when forget-objective losses are applied without constraint in other unlearning settings~\cite{zhangnegative}.

\vspace{0.5mm}
\noindent \textbf{Latent Space Crowding and Triplet Loss.} Triplet loss effectiveness also diminishes at scale, offering only marginal privacy gains over the single-speaker setting. As the forget set grows, pushing an embedding away from one negative sample in $\mathcal{F}$ inadvertently pushes it toward another~\cite{sohn2016improved}, since forget speakers now occupy overlapping regions of the representation space. Higher Max-FSSIM scores at 100 speakers confirm this limitation, underscoring that simultaneous suppression of multiple identities remains an open challenge for future work.

\section{Conclusion and Llimitation}
In this work, we formulate \emph{Speech Generation Speaker Poisoning} (SGSP) for zero-shot TTS systems and introduce parameter-modification baselines, Teacher-Guided Poisoning (TGP) and Encoder-Guided Poisoning (EGP), together with an comprehensive evaluation framework based on AUC and Forget Set Similarity (FSSIM) under easy and strong privacy conditions. For StyleTTS2 on LibriTTS-R, evaluated using a WavLM-based speaker verifier, our results show that targeted speaker suppression can be achieved under pairwise similarity metrics for up to 15 forget speakers, although Max-FSSIM indicates that worst-case identity leakage remains unresolved at larger scales. Under this setting, seen speakers are suppressed less effectively than unseen speakers, supporting the seen-speaker setting as a stricter benchmark. Experiments with 100 forget speakers further suggest scalability limitations associated with increasing overlap in the learned speaker representation space. These findings are based on a controlled evaluation using a single architecture, corpus, and speaker verifier, without subjective listening evaluation. Extending the study to additional architectures, datasets, independent verifiers, perceptual assessments, and adversarial recovery settings will help establish their broader generalizability.

\section{Acknowledgement}
This work was supported by the Office of the Director of National Intelligence (ODNI), Intelligence Advanced Research Projects Activity (IARPA), via the ARTS Program under contract D2023-2308110001. The views and conclusions contained herein are those of the authors and should not be interpreted as necessarily representing the official policies, either expressed or implied, of ODNI, IARPA, or the U.S. Government. The U.S. Government is authorized to reproduce and distribute reprints for governmental purposes notwithstanding any copyright annotation therein.

\section{Generative AI Use Disclosure}
Generative AI tools were used solely to assist with language refinement, including grammar correction, sentence restructuring, and improvements in clarity and readability. The use of these tools was limited to polishing the presentation of the manuscript and did not contribute to the development of the research. All research concepts, problem formulation, methodological design, experimental setup, analysis, and conclusions were developed by the authors. 

\bibliographystyle{IEEEtran}
\bibliography{mybib}

\end{document}